\documentclass[conference]{IEEEtran}
\usepackage{cite}
\usepackage{amsmath,amssymb,amsfonts}
\usepackage{algorithmic}
\usepackage{graphicx}
\usepackage{textcomp}
\usepackage{xcolor}
\usepackage{booktabs}
\usepackage{enumitem}
\usepackage{url}
\usepackage{tabularx}
\def\BibTeX{{\rm B\kern-.05em{\sc i\kern-.025em b}\kern-.08em
    T\kern-.1667em\lower.7ex\hbox{E}\kern-.125emX}}
\begin{document}

\title{Systematic Capability Benchmarking of Frontier Large Language Models for Offensive Cyber Tasks}

\author{ 
    \fontsize{10}{12}\selectfont
    Tyler H. Merves\textsuperscript{1},
    Michael H. Conaway\textsuperscript{1},
    Joseph M. Escobar\textsuperscript{1}, 
    Hakan T. Otal\textsuperscript{2},
    and Unal Tatar\textsuperscript{1} \\ 
    *\textit{Corresponding Author: utatar@albany.edu} \\ \\
    
    \begin{tabularx}{\textwidth}{>{\centering\arraybackslash}X}
        \fontsize{9}{10}\selectfont \textsuperscript{1} \textit{Department of Cybersecurity}  \\ 
        \fontsize{9}{10}\selectfont \textsuperscript{2} \textit{Department of Information Sciences and Technology} \\ 
        \fontsize{9}{10}\selectfont \textit{College of Emergency Preparedness, Homeland Security, and Cybersecurity} \\
        \fontsize{9}{10}\selectfont \textit{University at Albany, SUNY} \\
        \fontsize{9}{10}\selectfont \textit{Albany, NY, USA}
    \end{tabularx}
}

\maketitle

\begin{abstract}
We present, to our knowledge, the most comprehensive cross-model evaluation of LLM agents on offensive cybersecurity tasks, benchmarking 10 frontier models from 7 providers on all 200 challenges of the NYU CTF Bench. Building on the D-CIPHER multi-agent framework, we extend it with multi-provider backend support, a custom Kali Linux environment with over 100 pre-installed penetration testing tools, and runtime tool-discovery agents. Through a controlled factorial study, we find that the Kali Linux environment yields a $+9.5$ percentage-point improvement over Ubuntu, while auto-prompting and category-specific tips often degrade performance in well-equipped environments. Among models, Claude 4.5 Opus achieves the highest solve rate (59\%), followed by Gemini 3 Pro (52\%), with Gemini 3 Flash offering the best cost-efficiency at \$0.05 per solve. Asymmetric planner/executor model assignments provide no meaningful benefit while coherent same-model configurations consistently outperform mixed-tier pairings. Our results indicate that environment tooling and model selection emerge as the strongest drivers of performance, whereas prompt engineering interventions show diminishing or negative returns in well-equipped environments. Reported performance reflects both model reasoning ability and compatibility with agent tooling and API integration.
\end{abstract}


\section{Introduction}

Rapid advances in large language models (LLMs) have extended their capabilities into complex technical domains, including cybersecurity~\cite{ferrag2025generative}. LLM-powered agents can now discover vulnerabilities~\cite{fang2024llmagents}, craft exploits, and conduct multi-step penetration testing with limited human oversight~\cite{pentestgpt}. These capabilities raise dual-use concerns, as the same models that automate defensive workflows may also lower the barrier for offensive cyber operations~\cite{occult}. Systematic evaluation methodologies are therefore needed to quantify offensive cyber performance under controlled conditions.

Capture The Flag (CTF) competitions serve as a standardized proxy for measuring offensive security capability, requiring cryptanalysis, reverse engineering, binary exploitation, web application attacks, and forensic analysis within sandboxed environments. The NYU CTF Bench~\cite{nyuctfbench} formalizes this approach with 200 challenges spanning six categories. Building on this benchmark, increasingly sophisticated agent architectures have been proposed: EnIGMA~\cite{enigma} added interactive tool interfaces, D-CIPHER~\cite{dcipher} introduced a multi-agent Planner--Executor architecture, and CRAKEN~\cite{craken} augmented D-CIPHER with retrieval-augmented generation. Yet comparisons across these studies remain difficult because they differ in model selection, environment, prompt strategy, and evaluation protocol.

To our knowledge, no prior study systematically varies these engineering dimensions on the same benchmark under controlled conditions, making it unclear which factors truly drive performance. This paper addresses that gap through a comprehensive study on the full NYU CTF Bench. We extend D-CIPHER with multi-provider backend support, a custom Kali Linux environment with over 100 pre-installed security tools, and runtime tool-discovery agents. Our contributions are:

\begin{enumerate}
  \item A systematic comparison of 10 frontier LLMs from 7 providers on 200 CTF challenges, one of the most comprehensive cross-model evaluations on this benchmark to date.
  \item A controlled ablation across environment (Ubuntu vs.\ Kali), prompt strategy (generic vs.\ tips), and auto-prompting, showing that Kali Linux yields $+9.5$\,percentage-point (pp) improvement while auto-prompting generally hurts.
  \item A cost-performance analysis revealing that Claude 4.5 Opus achieves the highest solve rate (59\%) while Gemini 3 Flash offers the best cost-efficiency at \$0.05 per solve.
  \item Reproducible experimental infrastructure, including a Kali Linux Docker image with 100+ security tools, runtime tool-discovery agents, a parallel experiment runner, and an automated result parsing pipeline. Source code will be made available upon publication to enable replication and extension of our findings.
  \footnote{Source code available at: https://github.com/TATAR-LAB/ctf-agents}
\end{enumerate}

\section{Related Work}

\subsection{CTF Benchmarks}

Several benchmarks evaluate LLM capabilities on offensive security tasks. The NYU CTF Bench~\cite{nyuctfbench} provides 200 challenges from the annual CSAW CTF competitions (2016--2024) across six categories with Dockerized environments and ground-truth flags; in its initial evaluation, GPT-4 achieved only single-digit solve rates~\cite{shao2024empirical}. Cybench~\cite{cybench} offers 90 challenges from recent competitions and introduces subtask decomposition for partial-credit evaluation. CTFusion~\cite{ctfusion} demonstrates that data contamination inflates results on static benchmarks: an agent with web search nearly doubled its solve rate by retrieving published writeups, while performance on live competitions dropped by roughly half. Although absolute solve rates should be interpreted cautiously, relative comparisons between configurations are likely to remain valid, as all models are evaluated under identical conditions. BountyBench~\cite{bountybench} moves beyond CTFs to evaluate agents on real-world bug bounties, finding that defensive tasks are substantially easier than offensive ones. OCCULT~\cite{occult} complements these CTF-centric benchmarks by evaluating LLMs against realistic offensive cyber operation tactics used by modern threat actors, finding significant recent advancement in AI-enabled cyber risk. However, differences in training data exposure and tool integration across providers may still influence observed performance.

\subsection{LLM Agents for Offensive Security}

PentestGPT~\cite{pentestgpt} was among the first to apply LLMs to multi-step offensive security tasks. Its key finding that context loss over long interaction sequences is the primary bottleneck motivated subsequent multi-agent designs. EnIGMA~\cite{enigma} builds on SWE-agent with Interactive Agent Tools (IATs) for operating interactive programs such as debuggers and server connections, achieving 3$\times$ higher solve rates than the NYU CTF Bench baseline. EnIGMA also identified a ``soliloquizing'' phenomenon where agents hallucinate observations from memorized training data, and found that most successful solutions occur within the first 20 steps, which informs our timeout design.

D-CIPHER~\cite{dcipher} introduces the hierarchical Planner--Executor architecture that our work builds upon. By separating high-level strategy from low-level task execution, D-CIPHER achieved state-of-the-art results across NYU CTF Bench, Cybench, and HackTheBox, reporting a 6\% improvement over single-agent baselines and a 3\% gain from auto-prompting with Claude 3.5 Sonnet. CRAKEN~\cite{craken} extends D-CIPHER with retrieval-augmented generation over CTF writeups and attack payloads, achieving 22\% on NYU CTF Bench, notably the same as D-CIPHER, indicating that RAG augmentation did not improve performance on this benchmark.

Our work differs in scope: rather than proposing a new architectural component, we conduct the first controlled evaluation that systematically varies environment, prompting, model selection, and architecture assignment across 10 frontier LLMs on the full NYU CTF Bench.

\section{Methodology}

We extend the D-CIPHER multi-agent framework~\cite{dcipher} with new model backends, execution environments, and tooling, then conduct a systematic evaluation across four research questions on the NYU CTF Bench~\cite{nyuctfbench}.

\subsection{D-CIPHER Framework}

D-CIPHER is a multi-agent system built on a hierarchical Planner--Executor architecture designed for solving CTF challenges~\cite{dcipher}. Mirroring collaborative CTF teams, a \textit{Planner} agent handles high-level problem decomposition, reasoning about attack strategies, and delegating sub-tasks to one or more \textit{Executor} agents. Each Executor carries out specific low-level operations (running shell commands, invoking security tools, generating and executing code, and interpreting outputs) and reports results back to the Planner for iterative refinement. This separation of concerns breaks complex challenges into manageable steps, increasing solving efficiency over monolithic single-agent approaches.

D-CIPHER also includes an optional \textit{Auto-Prompter} agent that runs before the Planner--Executor loop. The Auto-Prompter serves as a foundational step to increase the likelihood that the system follows a productive solution path. It explores the challenge's Docker environment (examining file structures, identifying available tools, and reading challenge descriptions), then generates a tailored initial prompt intended to steer the system toward productive attack paths. The original D-CIPHER study reported that auto-prompting improved solve rates by approximately 3\% with Claude 3.5 Sonnet~\cite{dcipher}; however, as we discuss in Section~IV, our experiments with Gemini 3 Pro yield different conclusions about its efficacy.

Additionally, D-CIPHER leverages the categorical structure of the NYU CTF Bench by implementing specialized \textit{tips} for each challenge category. These tips provide actionable guidance for approaching challenges using common attack paths of each category, such as suggesting the use of \texttt{pwntools} for binary exploitation or recommending frequency analysis for cryptography challenges.

\begin{figure}[htb]
  \centering
  \includegraphics[width=\columnwidth]{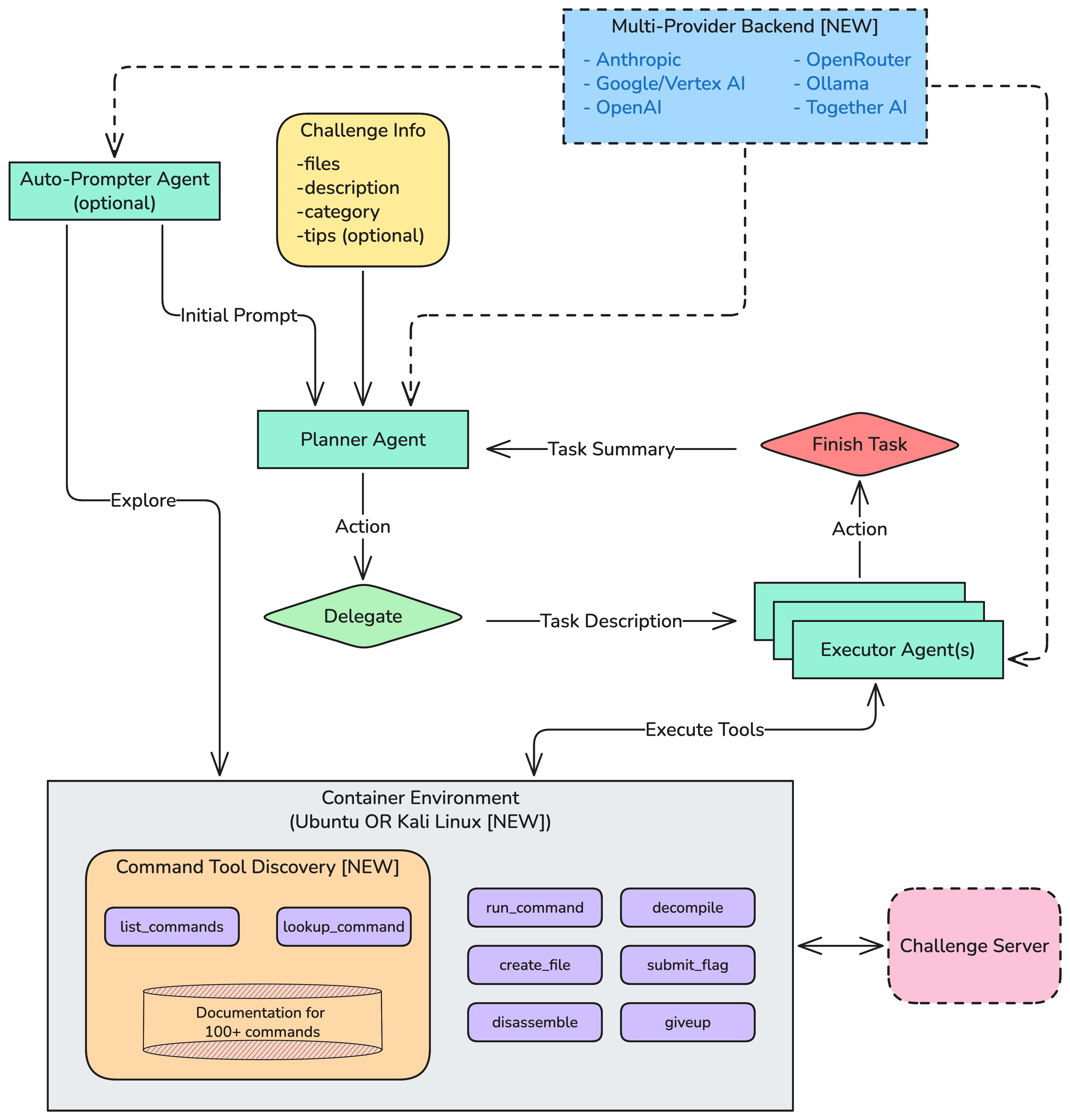}
  % \fbox{\parbox{0.9\columnwidth}{\centering\vspace{3cm}\textit{Architecture diagram placeholder}\vspace{3cm}}}
  \caption{Overview of the extended D-CIPHER multi-agent architecture.}
  \label{fig:architecture}
\end{figure}

\smallskip
\noindent\textbf{Extensions to D-CIPHER.}
The original D-CIPHER framework utilizes an Ubuntu environment with a limited amount of pre-installed cybersecurity tools and supports only OpenAI-formatted API endpoints, limiting both tooling and model selection. We introduce three major extensions (Figure~\ref{fig:architecture}):

\begin{enumerate}[leftmargin=*]
  \item \textbf{Multi-provider backend support.} We added Vertex AI, OpenRouter, and Ollama backends, increasing the number of supported model providers. This enables systematic cross-model comparison across proprietary and open-source LLMs within a unified framework.
  
  \item \textbf{Kali Linux execution environment.} We built a custom Docker image based on Kali Linux with over 100 pre-installed penetration testing tools (e.g., \texttt{nmap}, \texttt{sqlmap}, \texttt{john}, \texttt{hashcat}, \texttt{binwalk}, \texttt{volatility}) and a structured documentation file (\texttt{commands\_documentation.csv}) containing usage descriptions for all available commands.
  
  \item \textbf{Tool discovery agents.} To ensure agents can leverage the expanded Kali toolset, we created two new tools that are available exclusively in the Kali environment: \texttt{ListCommandsTool}, which enumerates all available commands with brief descriptions, and \texttt{LookupCommandTool}, which queries compiled documentation for individual commands. Together, these tools reduce tool-discovery friction by allowing agents to first identify relevant commands and then verify their usage before invocation. The performance improvement observed with the Kali environment (Section~IV-A) reflects the combined effect of this expanded toolset and the tool discovery capabilities.
\end{enumerate}

\subsection{Benchmark Dataset}

We evaluate on the NYU CTF Bench~\cite{nyuctfbench}, a dataset of 200 Capture The Flag challenges sourced from the Cyber Security Awareness Week (CSAW) CTF competitions hosted annually at New York University from 2016 to 2024. The challenges span six categories: cryptography~(52), reverse engineering~(51), binary exploitation~(39), miscellaneous~(24), web~(19), and forensics~(15). Difficulty ranges from beginner to advanced, reflecting a broad spectrum of offensive security tasks encountered by cybersecurity professionals, from web application attacks to low-level binary exploitation. Each challenge includes source files for initialization, Docker configurations for reproducible environmental setup, and a ground-truth flag that enables automated verification of candidate solutions.

\subsection{Experimental Design}

We address four research questions through controlled experiments on the full 200-challenge set. Unless otherwise noted, experiments use Gemini 3 Pro (accessed via Vertex AI) as the default model. All agents share the following hyperparameters: temperature $T{=}1.0$ (following the D-CIPHER recommendation for optimal performance~\cite{dcipher}), 30 planner rounds (100 executor rounds), a nominal \$5.00 per-challenge cost limit (never reached in practice), and a 10-minute wall-clock timeout. The timeout reflects findings from D-CIPHER~\cite{dcipher} and EnIGMA~\cite{enigma} that most successful solves complete within the first few minutes, with longer runs exhibiting diminishing returns and predominantly ending in failure; a 10-minute cutoff therefore captures the empirically observed productive window while keeping costs tractable across thousands of runs. All experiments report the following metrics: solve rate (number of challenges solved out of 200), total API cost, cost per solved challenge, and per-category solve rates.

\subsubsection{RQ1 \& RQ2 (Environment and Prompt Engineering)}

RQ1 and RQ2 both address the effectiveness of the D-CIPHER framework with varying configurations.

\noindent\textit{RQ1: Does enhanced security tooling improve LLM performance on CTF challenges?} We hypothesize that the richer tooling and documentation-retrieval capabilities of the Kali environment (Section~III-A) increase solve rates by reducing tool-discovery friction and enabling more sophisticated attack strategies.

\noindent\textit{RQ2: How do prompting strategies affect performance across different environments?} We examine whether category-specific tips and auto-prompting generalize to the Kali Linux environment, given that D-CIPHER's original 3\% gain from tips was observed with Claude 3.5 Sonnet on Ubuntu~\cite{dcipher}.

To address both questions jointly, we conduct a controlled ablation across three binary factors:

\begin{itemize}[leftmargin=*]
  \item \textbf{Environment}: Ubuntu 22.04 (base Docker image with standard tools such as \texttt{gdb}, \texttt{radare2}, \texttt{pwntools}, and \texttt{angr}) versus Kali Linux (custom Docker image with 100+ penetration testing tools and runtime-queryable documentation).
  \item \textbf{Prompt strategy}: Generic prompts (minimal task instructions) versus category-specific tips (actionable guidance tailored to each challenge category).
  \item \textbf{Auto-prompting}: Off (planner receives only the static prompt) versus On (auto-prompter explores challenge files and generates a tailored initial prompt before the solve attempt begins).
\end{itemize}

\noindent The combination of these three binary factors yields eight configurations, each systematically run on all 200 challenges with Gemini 3 Pro, for a total of 1,600 challenge runs.

\subsubsection{RQ3 (Model Selection)}

\noindent\textit{Which LLMs perform best on offensive cyber tasks?} We hypothesize that larger frontier models tend to outperform smaller and open-source models due to superior tool-use, multi-turn reasoning, and instruction-following capabilities.

Using the best configuration identified in RQ1/RQ2 (Kali + generic prompts, no auto-prompting), we compared ten models from seven providers, spanning several specialization types:

\begin{itemize}[leftmargin=*]
  \item \textbf{Frontier:} Claude 4.5 Opus (Anthropic), Gemini 3 Pro (Google), GPT-5.2 (OpenAI)
  \item \textbf{Provider-flagship:} GLM-5 (Z-AI)
  \item \textbf{Small/efficient:} Gemini 3 Flash (Google)
  \item \textbf{Code-specialized:} GPT-5.2-Codex (OpenAI)
  \item \textbf{Reasoning:} DeepSeek-R1 (DeepSeek)
  \item \textbf{Open-source:} DeepSeek-V3 (DeepSeek), Llama 3.3 70B (Meta), Qwen 3.5 397B-A17B (Alibaba)
\end{itemize}

\noindent To ensure a fair comparison, all models share the identical hyperparameters, toolset, and timeout described above.

\subsubsection{RQ4 (Multi-Agent Architecture)}

\noindent\textit{Does asymmetric model assignment across planner and executor roles improve cost-performance over symmetric configurations?} We hypothesize that using a larger, more capable model for the Planner and a smaller, faster model for the Executor would achieve better cost-performance balance than a uniform model assignment. This assumes that planning demands high-level strategic reasoning, whereas execution is more mechanical, involving tool calls and command translation.

We assessed two symmetric configurations (Gemini 3 Pro for both roles, and Gemini 3 Flash for both roles) and two asymmetric configurations (Pro Planner + Flash Executor, and Flash Planner + Pro Executor), all under the Kali + generic prompt setup.

\section{Experimental Results}

We present results for each of the four research questions on the full 200-challenge NYU CTF Bench.

\subsection{Impact of Environment and Prompt Strategy}

\begin{figure}[htb]
  \centering
  \includegraphics[width=\columnwidth]{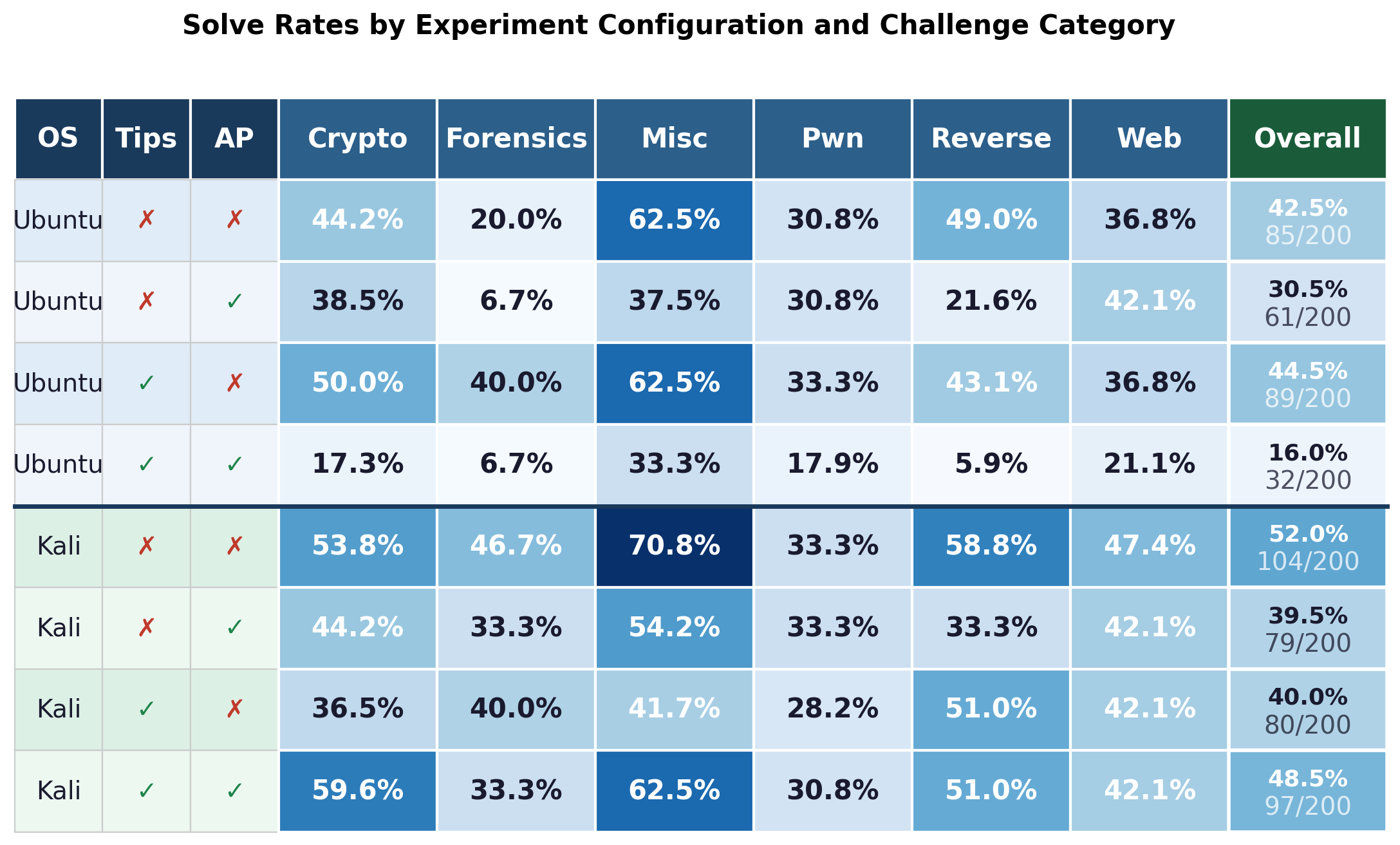}
  \caption{Solve rates by configuration and challenge category (RQ1+RQ2).}
  \label{fig:rq12_heatmap}
\end{figure}

Figure~\ref{fig:rq12_heatmap} summarizes the results across all eight configurations (Environment $\times$ Prompts $\times$ AutoPrompt), all run with Gemini 3 Pro.

% \noindent\textbf{Environment effect (RQ1).}
Kali Linux consistently outperforms Ubuntu under matched conditions. Under identical prompt settings (Generic, no AutoPrompt), Kali achieves 52.0\% (104/200) versus 42.5\% (85/200) for Ubuntu, a gain of $+9.5$\,pp. The advantage is especially pronounced in forensics ($46.7\%$ vs.\ $20.0\%$) and reverse engineering ($58.8\%$ vs.\ $49.0\%$), categories that depend heavily on specialized tool availability.

% \noindent\textbf{Prompt strategy effect (RQ2).}
The interaction between environment and prompt strategy is noteworthy. Under Kali, generic prompts outperform category-specific tips ($52.0\%$ vs.\ $40.0\%$), whereas under Ubuntu the pattern reverses ($42.5\%$ vs.\ $44.5\%$). One possible explanation is that the richly tooled Kali environment already encodes implicit guidance; layering explicit tips over-constrains the agent's exploration. On Ubuntu, the sparser environment means tips compensate by steering the agent toward appropriate techniques.

% \noindent\textbf{Auto-prompting effect.}
Auto-prompting, in which the agent explores challenge files and generates a tailored prompt before solving, generally reduces performance across most configurations. Under Kali + Generic, AutoPrompt reduces the solve rate from $52.0\%$ to $39.5\%$ ($-12.5$\,pp); under Ubuntu + Tips, the drop is even sharper: $44.5\%$ to $16.0\%$ ($-28.5$\,pp). One exception stands out: under Kali + Tips, AutoPrompt raises the rate from $40.0\%$ to $48.5\%$ ($+8.5$\,pp), suggesting it can partially recover a sub-optimal base strategy. We hypothesize that the additional exploration phase typically consumes budget without producing actionable insight, and may introduce misleading preliminary analysis.

The best overall configuration (Kali + Generic without AutoPrompt) costs \$44.23 total. The most expensive configuration (Kali + Tips + AutoPrompt, \$66.12) achieves only the second-best solve rate (48.5\%), reinforcing that simpler configurations can be more effective. Since this ablation uses only Gemini 3 Pro, the magnitude of these effects may vary with other models; however, the directional findings (environment over prompting) are consistent with the broader patterns observed in RQ3.

\subsection{Model Selection}

\begin{figure}[htb]
  \centering
  \includegraphics[width=\columnwidth]{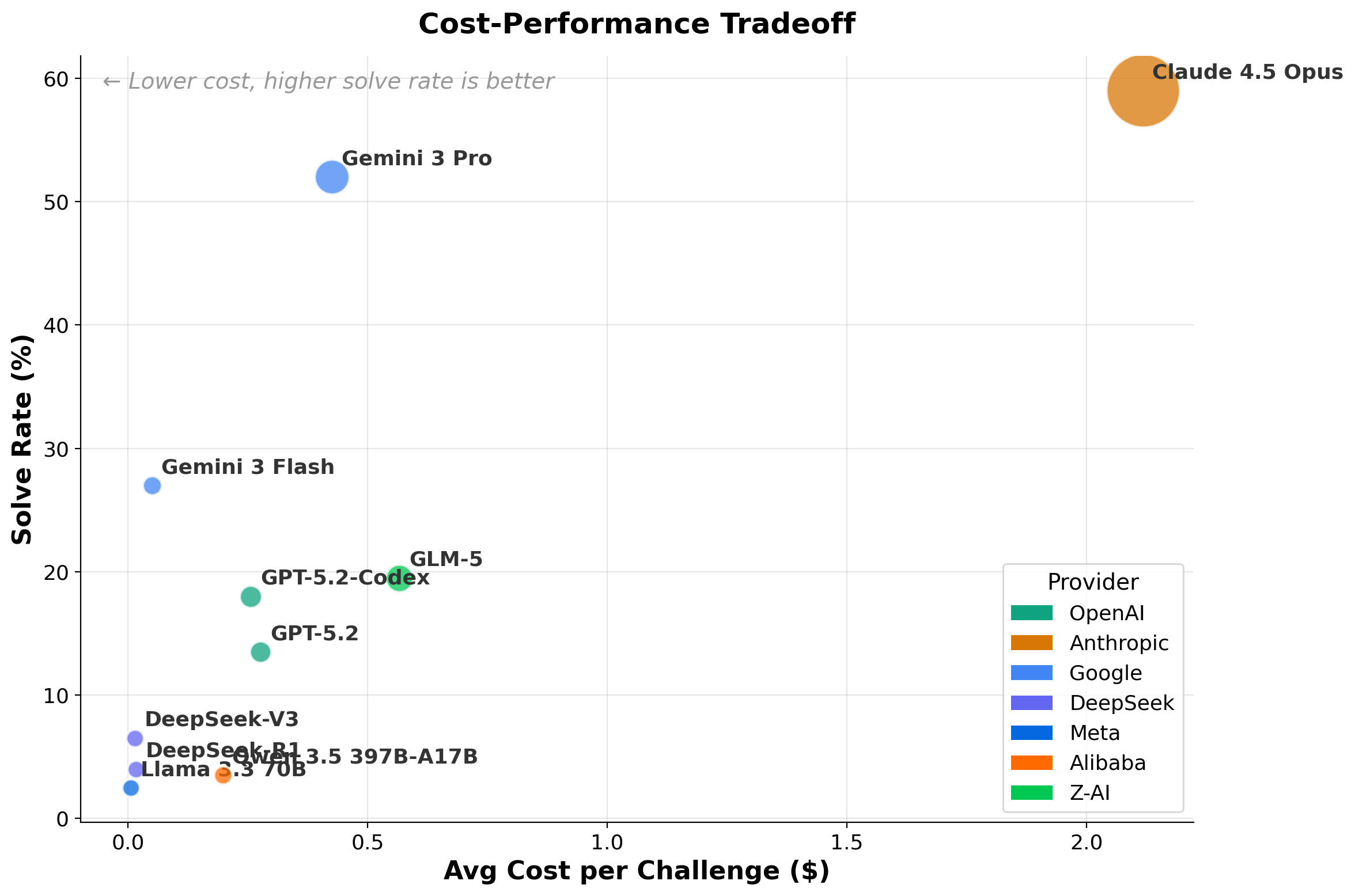}
  \caption{Cost--performance tradeoff: average cost per challenge vs.\ solve rate. Bubble size is proportional to total cost.}
  \label{fig:rq3_cost}
\end{figure}

We benchmark ten LLMs from seven providers under the best RQ1/RQ2 configuration (Kali + Generic, no AutoPrompt). Table~\ref{tab:rq3_results} reports solve rates, total costs, and cost-per-solve; Figure~\ref{fig:rq3_cost} visualizes the cost--performance tradeoff.

\begin{table}[htb]
\centering
\caption{Model benchmark on NYU CTF Bench \\(200 Challenges on Kali with Generic prompts)}
\label{tab:rq3_results}
\setlength{\tabcolsep}{4pt}
\footnotesize
\begin{tabular}{@{}llrrrr@{}}
\toprule
\textbf{Model} & \textbf{Provider} & \textbf{Solv.} & \textbf{Rate} & \textbf{Cost} & \textbf{\$/Solv.} \\
\midrule
Claude 4.5 Opus    & Anthropic & 118 & 59.0\% & 249.92 & 2.12 \\
Gemini 3 Pro       & Google    & 104 & 52.0\% &  44.23 & 0.43 \\
Gemini 3 Flash     & Google    &  54 & 27.0\% &   2.69 & 0.05 \\
GLM-5              & Z-AI      &  39 & 19.5\% &  22.09 & 0.57 \\
GPT-5.2-Codex      & OpenAI    &  36 & 18.0\% &   9.23 & 0.26 \\
GPT-5.2            & OpenAI    &  27 & 13.5\% &   7.47 & 0.28 \\
DeepSeek-V3        & DeepSeek  &  13 &  6.5\% &   0.19 & 0.01 \\
DeepSeek-R1        & DeepSeek  &   8 &  4.0\% &   0.13 & 0.02 \\
Qwen 3.5 397B-A17B & Alibaba   &   7 &  3.5\% &   1.39 & 0.20 \\
Llama 3.3 70B      & Meta      &   5 &  2.5\% &   0.03 & 0.01 \\
\bottomrule
\end{tabular}
\end{table}

For context, prior state-of-the-art results on NYU CTF Bench were 22\% by D-CIPHER~\cite{dcipher} and 22\% by CRAKEN~\cite{craken}, both using Claude 3.5 Sonnet, while the original benchmark evaluation with GPT-4 achieved only single-digit solve rates~\cite{nyuctfbench}. Our best configurations exceed these baselines; however, direct comparison is limited by generational model differences. Notably, even our Ubuntu baseline with Gemini 3 Pro (42.5\%) exceeds prior results, suggesting that the improvement is driven primarily by advances in model capability rather than our framework extensions alone.

% \noindent\textbf{Overall ranking.}
Claude 4.5 Opus leads at 59.0\% (118/200), followed by Gemini 3 Pro at 52.0\% (104/200). These two frontier models form a clear top tier, each solving nearly twice as many challenges as the third-ranked Gemini 3 Flash (27.0\%). OpenAI's models cluster in the middle tier: GPT-5.2-Codex (18.0\%) outperforms base GPT-5.2 (13.5\%), suggesting code specialization provides a meaningful advantage. Open-source and smaller models achieve lower solve rates: GLM-5 (19.5\%) attains moderate accuracy but at higher cost (\$0.57/solve), while DeepSeek-V3 (6.5\%), DeepSeek-R1 (4.0\%), Qwen 3.5 (3.5\%), and Llama 3.3 70B (2.5\%) all fall below 7\%. These lower results may reflect a combination of weaker tool-use capabilities, less robust multi-turn agentic reasoning, and differences in how well each model integrates with the agent framework's tool-calling interface.

% \noindent\textbf{Reasoning vs.\ agentic capability.}
The DeepSeek pair provides an instructive comparison: the reasoning-specialized R1 (4.0\%) does not outperform the standard V3 (6.5\%). While both models achieve low solve rates overall, this pattern is consistent with the observation that agentic settings, where success depends on iterative tool calls and environment interaction, may reward robust tool-use over pure chain-of-thought reasoning. However, both models' low scores make it difficult to draw strong conclusions from this comparison alone.

\begin{figure}[htb]
  \centering
  \includegraphics[width=\columnwidth]{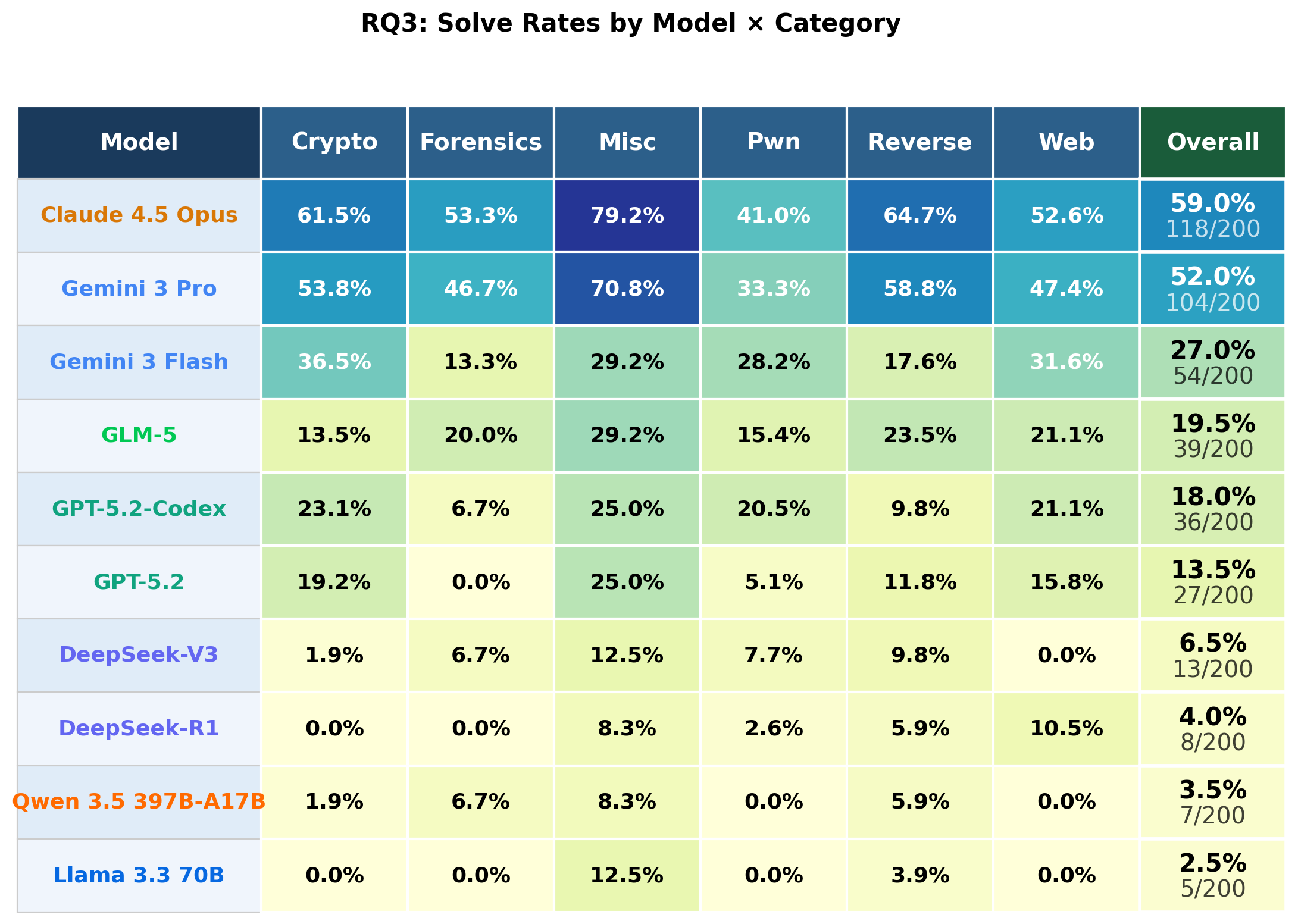}
  \caption{Per-category solve rates across all ten models. Darker cells indicate higher solve rates.}
  \label{fig:rq3_heatmap}
\end{figure}

% \noindent\textbf{Category patterns.}
Per-category analysis (Figure~\ref{fig:rq3_heatmap}) reveals that the \textit{misc} category is the easiest across nearly all models, while \textit{pwn} is consistently the hardest. Miscellaneous challenges often involve pattern recognition, encoding, or scripting tasks that align well with LLM strengths, whereas binary exploitation requires precise memory layout reasoning and multi-step exploit construction that current models struggle to sustain. Claude 4.5 Opus outperforms Gemini 3 Pro in every category, with the largest gains in pwn (41.0\% vs.\ 33.3\%) and web (52.6\% vs.\ 47.4\%).

% \noindent\textbf{Cost-efficiency.}
Gemini 3 Flash occupies a favorable position on the cost--performance frontier (Figure~\ref{fig:rq3_cost}): at \$0.05/solve, it is the most cost-efficient model tested, solving over half of what Pro achieves at 6\% of the cost. Claude 4.5 Opus achieves the highest solve rate but at the highest total cost (\$249.92, \$2.12/solve), roughly 5$\times$ more expensive per solve than Gemini 3 Pro (\$44.23, \$0.43/solve). For cost-sensitive deployments, Gemini 3 Flash offers the best tradeoff, while Claude 4.5 Opus is the best option when accuracy is the primary objective regardless of cost.

\subsection{Multi-Agent Architecture}

Table~\ref{tab:rq4_results} presents results for four planner/executor configurations using Gemini 3 Pro and Flash, all under Kali + Generic.

\begin{table}[htb]
\centering
\caption{Planner/Executor architecture comparison \\(200 challenges, Kali + Generic).}
\label{tab:rq4_results}
\footnotesize
\begin{tabular}{@{}llrrr@{}}
\toprule
\textbf{Planner} & \textbf{Executor} & \textbf{Solved} & \textbf{Rate} & \textbf{Cost (\$)} \\
\midrule
Pro   & Pro   & 104 & 52.0\% & \$44.23 \\
Pro   & Flash &  57 & 28.5\% &  \$3.90 \\
Flash & Flash &  54 & 27.0\% &  \$2.69 \\
Flash & Pro   &  47 & 23.5\% &  \$2.92 \\
\bottomrule
\end{tabular}
\end{table}

The homogeneous Pro configuration achieves 52.0\%, nearly doubling the next-best result. The asymmetric Pro Planner + Flash Executor yields only $+1.5$\,pp over Flash-only (28.5\% vs.\ 27.0\%) while increasing cost by 45\%.

Notably, the reverse configuration (Flash Planner + Pro Executor) performs worse than Flash-only (23.5\% vs.\ 27.0\%), despite deploying the stronger model as executor. This challenges the intuition that executor quality is the bottleneck; instead, it suggests that \textbf{coherent planning and execution within the same model family and capability tier} matters more than the raw capability of either role alone. We hypothesize that mismatched models introduce coordination inefficiencies between planning and execution: the planner's instructions may not align with how the executor model interprets and decomposes tasks, leading to wasted rounds and miscoordination. A weaker planner may issue underspecified or naive instructions that a stronger executor cannot compensate for, whereas a same-model pair shares implicit assumptions about instruction granularity and tool-use patterns. The performance gap is largest in categories requiring sustained multi-step reasoning: crypto (53.8\% for Pro-only vs.\ 21.1\% for Pro+Flash) and misc (70.8\% vs.\ 29.2\%). The practical implication is that using the same model for both roles appears more effective than mixing model tiers. We note that these findings are based on two models from the same provider (Google); whether the coherence effect generalizes across providers remains an open question.

\section{Conclusion}

This paper presents a systematic evaluation of frontier large language model (LLM) agents on offensive cybersecurity tasks using the NYU CTF Bench. By varying environment, prompting strategy, model selection, and multi-agent architecture, we identify the primary drivers of performance in agent-based cyber operations.

Our results show that environment and tooling are important: a Kali Linux setup with pre-installed security tools improves performance over a standard Ubuntu environment. In contrast, prompt engineering techniques such as category-specific tips and auto-prompting do not consistently improve performance and can degrade results in well-equipped environments. Model selection remains a dominant factor, with frontier models outperforming smaller and open-source alternatives. We also find that homogeneous planner–executor configurations outperform mixed-model setups, highlighting the importance of coordination within a single model.

These findings suggest that environment configuration and model choice have greater impact than additional prompt complexity, while also emphasizing that performance depends not only on model capability but also on tool integration and API interaction.

This study has several limitations. Results are specific to the D-CIPHER framework, and other architectures may exhibit different behavior. Because LLMs are inherently nondeterministic and all experiments use a sampling temperature of $T{=}1.0$, solve rates may vary across runs; reported figures represent single-trial measurements and should be interpreted as indicative rather than exact. The NYU CTF Bench reflects academic CTF tasks, and model rankings correspond to early-2026 versions. Additionally, as shown in prior work, static CTF benchmarks are susceptible to data contamination. While absolute solve rates may be inflated, relative comparisons remain meaningful due to shared exposure. Validation on live or dynamic benchmarks would strengthen external validity.

Future work will extend this evaluation to additional benchmarks such as CyBench, explore retrieval-augmented generation and adaptive task routing, and investigate persistent challenges in categories like binary exploitation and forensics. Developing standardized and reproducible benchmarking protocols is also an important direction.

As LLM capabilities in offensive cybersecurity continue to advance, rigorous and transparent evaluation will be essential for both capability assessment and responsible use.

% \section*{Ethical Considerations}

% All experiments were conducted within sandboxed Docker containers against synthetic CTF challenges with no connection to live systems, networks, or real-world targets. The benchmark is used strictly to measure and compare LLM capabilities under controlled conditions, not to develop or distribute offensive tools. We recognize the dual-use nature of this research and release our infrastructure to support defensive evaluation and AI safety research rather than to lower the barrier for malicious use.

\section*{Acknowledgment}

This work was supported by the SUNY AI Platform, powered by Google Cloud, and the Griffiss Institute (Award Number SA1001202200481). The contents of this paper are the sole responsibility of the authors and do not necessarily represent the official views of the sponsors.

\bibliography{paper}

@misc{dcipher,
  title     = {{D-CIPHER}: Dynamic Collaborative Intelligent Multi-Agent System with Planner and Heterogeneous Executors for Offensive Security},
  author    = {Udeshi, Meet and Shao, Minghao and Xi, Haoran and Rani, Nanda and Milner, Kimberly and Putrevu, Venkata Sai Charan and Dolan-Gavitt, Brendan and Shukla, Sandeep Kumar and Krishnamurthy, Prashanth and Khorrami, Farshad and Karri, Ramesh and Shafique, Muhammad},
  year      = {2025},
  url       = {http://arxiv.org/abs/2502.10931},
  doi       = {10.48550/arXiv.2502.10931},
  note      = {arXiv:2502.10931}
}

@misc{nyuctfbench,
  title     = {{NYU CTF Bench}: A Scalable Open-Source Benchmark Dataset for Evaluating {LLMs} in Offensive Security},
  author    = {Shao, Minghao and Jancheska, Sofija and Udeshi, Meet and Dolan-Gavitt, Brendan and Xi, Haoran and Milner, Kimberly and Chen, Boyuan and Yin, Max and Garg, Siddharth and Krishnamurthy, Prashanth and Khorrami, Farshad and Karri, Ramesh and Shafique, Muhammad},
  year      = {2025},
  url       = {http://arxiv.org/abs/2406.05590},
  doi       = {10.48550/arXiv.2406.05590},
  note      = {arXiv:2406.05590}
}

@misc{cybench,
  title     = {{Cybench}: A Framework for Evaluating Cybersecurity Capabilities and Risks of Language Models},
  author    = {Zhang, Andy K. and Perry, Neil and Dulepet, Riya and Ji, Joey and Menders, Celeste and Lin, Justin W. and Jones, Eliot and Hussein, Gashon and Liu, Samantha and Jasper, Donovan and others},
  year      = {2025},
  url       = {http://arxiv.org/abs/2408.08926},
  doi       = {10.48550/arXiv.2408.08926},
  note      = {arXiv:2408.08926}
}

@misc{enigma,
  title     = {{EnIGMA}: Interactive Tools Substantially Assist {LM} Agents in Finding Security Vulnerabilities},
  author    = {Abramovich, Talor and Udeshi, Meet and Shao, Minghao and Lieret, Kilian and Xi, Haoran and Milner, Kimberly and Jancheska, Sofija and Yang, John and Jimenez, Carlos E. and Khorrami, Farshad and Krishnamurthy, Prashanth and Dolan-Gavitt, Brendan and Shafique, Muhammad and Narasimhan, Karthik and Karri, Ramesh and Press, Ofir},
  year      = {2025},
  url       = {http://arxiv.org/abs/2409.16165},
  doi       = {10.48550/arXiv.2409.16165},
  note      = {arXiv:2409.16165}
}

@article{fang2024llmagents,
  title   = {{LLM} Agents can Autonomously Hack Websites},
  author  = {Fang, Richard and Bindu, Rohan and Gupta, Akul and Zhan, Qiusi and Kang, Daniel},
  journal = {arXiv preprint arXiv:2402.06664},
  year    = {2024}
}

@inproceedings{pentestgpt,
  title     = {{PentestGPT}: Evaluating and Harnessing Large Language Models for Automated Penetration Testing},
  author    = {Deng, Gelei and Liu, Yi and Mayoral-Vilches, V{\'\i}ctor and Liu, Peng and Li, Yuekang and Xu, Yuan and Zhang, Tianwei and Liu, Yang and Pinz{\'o}n, Martin and Rass, Stefan},
  booktitle = {USENIX Security Symposium},
  year      = {2024}
}

@misc{bountybench,
  title     = {{BountyBench}: Dollar Impact of {AI} Agent Attackers and Defenders on Real-World Cybersecurity Systems},
  author    = {Zhang, Andy K. and Ji, Joey and Menders, Celeste and Dulepet, Riya and Qin, Thomas and Wang, Ron Y. and Wu, Junrong and Liao, Kyleen and Li, Jiliang and Hu, Jinghan and others},
  year      = {2025},
  url       = {http://arxiv.org/abs/2505.15216},
  doi       = {10.48550/arXiv.2505.15216},
  note      = {arXiv:2505.15216}
}

@article{ctfusion,
  title   = {{CTFusion}: A {CTF}-Based Benchmark for {LLM} Agent Evaluation},
  author  = {Anonymous},
  journal = {Under review at ICLR},
  year    = {2026}
}

@misc{craken,
  title     = {{CRAKEN}: Cybersecurity {LLM} Agent with Knowledge-Based Execution},
  author    = {Shao, Minghao and Xi, Haoran and Rani, Nanda and Udeshi, Meet and Putrevu, Venkata Sai Charan and Milner, Kimberly and Dolan-Gavitt, Brendan and Shukla, Sandeep Kumar and Krishnamurthy, Prashanth and Khorrami, Farshad and Karri, Ramesh and Shafique, Muhammad},
  year      = {2025},
  url       = {http://arxiv.org/abs/2505.17107},
  doi       = {10.48550/arXiv.2505.17107},
  note      = {arXiv:2505.17107}
}

@misc{shao2024empirical,
  title     = {An Empirical Evaluation of {LLMs} for Solving Offensive Security Challenges},
  author    = {Shao, Minghao and Chen, Boyuan and Jancheska, Sofija and Dolan-Gavitt, Brendan and Garg, Siddharth and Karri, Ramesh and Shafique, Muhammad},
  year      = {2024},
  url       = {http://arxiv.org/abs/2402.11814},
  doi       = {10.48550/arXiv.2402.11814},
  note      = {arXiv:2402.11814}
}

@article{ferrag2025generative,
  title   = {Generative {AI} in Cybersecurity: A Comprehensive Review of {LLM} Applications and Vulnerabilities},
  author  = {Ferrag, Mohamed Amine and Alwahedi, Fatima and Battah, Ammar and Cherif, Bilel and Mechri, Abdechakour and Tihanyi, Norbert and Bisztray, Tamas and Debbah, Merouane},
  journal = {Internet of Things and Cyber-Physical Systems},
  year    = {2025},
  doi     = {10.1016/j.iotcps.2025.01.001}
}

@misc{occult,
  title     = {{OCCULT}: Evaluating Large Language Models for Offensive Cyber Operation Capabilities},
  author    = {Kouremetis, Michael and Dotter, Marissa and Byrne, Alex and Martin, Dan and Michalak, Ethan and Russo, Gianpaolo and Threet, Michael and Zarrella, Guido},
  year      = {2025},
  url       = {http://arxiv.org/abs/2502.15797},
  doi       = {10.48550/arXiv.2502.15797},
  note      = {arXiv:2502.15797}
}

\end{document}